\documentclass[amssymb, amsmath, reprint, prl, showpacs]{revtex4-1}

\usepackage{microtype}
\usepackage{graphicx}
\usepackage{xcolor}
\usepackage{siunitx}
\usepackage{hyperref}
\usepackage{cleveref}
\usepackage{braket}
\usepackage{color}
\usepackage{bbm}
\usepackage{pbox}
\usepackage{relsize}
\usepackage{diagbox}
\usepackage{enumerate}
\usepackage[english]{babel}

\usepackage{tikz}
\usepackage{tikz-qtree}

\definecolor{niceblue}{HTML}{337ab7}
\definecolor{nicegreen}{HTML}{5cb85c}
\definecolor{nicered}{HTML}{d9534e}

\DeclareMathOperator{\tr}{tr}

\newcommand{\ketbra}[2]{|#1\rangle\langle#2|}

\begin{document}

\title{No Fine theorem for macrorealism: Limitations of the Leggett-Garg inequality}
\date{\today}

\author{Lucas Clemente}
\email{lucas.clemente@mpq.mpg.de}
\author{Johannes Kofler}
\email{johannes.kofler@mpq.mpg.de}
\affiliation{Max Planck Institute of Quantum Optics, Hans-Kopfermann-Str.\ 1, 85748 Garching, Germany}

\begin{abstract}
  Tests of local realism and macrorealism have historically been discussed in very similar terms: Leggett-Garg inequalities follow Bell inequalities as necessary conditions for classical behavior. Here, we compare the probability polytopes spanned by all measurable probability distributions for both scenarios and show that their structure differs strongly between spatially and temporally separated measurements. We arrive at the conclusion that, in contrast to tests of local realism where Bell inequalities form a necessary and sufficient set of conditions, no set of inequalities can ever be necessary and sufficient for a macrorealistic description. Fine's famous proof that Bell inequalities are necessary and sufficient for the existence of a local realistic model, therefore cannot be transferred to macrorealism. A recently proposed condition, no-signaling in time, fulfills this criterion, and we show why it is better suited for future experimental tests and theoretical studies of macrorealism. Our work thereby identifies a major difference between the mathematical structures of local realism and macrorealism.
\end{abstract}

\pacs{03.65.Ta, 03.65.Ud}

\maketitle

The violation of classical world views, such as local realism \cite{Bell:1964wu} and macrorealism \cite{Leggett:1985bl, Leggett:2002dk}, is one of the most interesting properties of quantum mechanics. Experiments performed over the past decades have shown violations of local realism in various systems \cite{Freedman:1972ka, Aspect:1982ja, Weihs:1998cc}, while violations of macrorealism are on the horizon \cite{PalaciosLaloy:2010ih, Goggin:2011iw, Xu:2011cr, Dressel:2011hh, Fedrizzi:2011ji, Waldherr:2011km, Athalye:2011bi, Souza:2011hu, Zhou:2015fi, Knee:2012cg, Suzuki:2012dw, George:2013bd, Katiyar:2013dt, Emary:2014ck, Asadian:2014fw, Robens:2015baa, White:2015ui-arxiv, Knee:2016wd-arxiv}. The latter endeavors pave the way towards the experimental realization of Schr\"odinger's famous thought experiment \cite{Schrodinger:1935kq}. In the future, they might offer insight into important foundational questions, such as the quantum measurement problem \cite{Leggett:2005bf}, and allow experimental tests of (possibly gravitational) extensions of quantum mechanics \cite{RomeroIsart:2011es}.

Historically, the discussion of tests of macrorealism (MR) follows the discussion of tests of local realism (LR) closely: Leggett-Garg inequalities (LGIs) \cite{Leggett:1985bl} are formulated similarly to Bell inequalities \cite{Bell:1964wu, Clauser:1969ff, Clauser:1974fv}, and some concepts, e.g.\ quantum contextuality \cite{Kochen:1967vo}, are connected to both fields \cite{Avis:2010jm, Kleinmann:2012wr, Araujo:2013fq, Kujala:2015gk, Dzhafarov:2015ic}. However, recently, a discrepancy between LR and MR has been identified: Whereas Fine's theorem states that Bell \emph{inequalities} are both necessary and sufficient for LR \cite{Fine:1982ic}, a combination of arrow of time (AoT) and no-signaling in time (NSIT) \cite{Kofler:2013hb} \emph{equalities} are necessary and sufficient for the existence of a macrorealistic description \cite{Clemente:2015hv}. A previous study \cite{Clemente:2015hv} also demonstrated that LGIs involving temporal correlation functions of pairs of measurements are not sufficient for macrorealism, but did not rule out a potential sufficiency of other sets of LGIs, e.g.\ of the CH type \cite{Clauser:1974fv, Mal:2015wn-arxiv}, leaving open the possibility of a Fine theorem for macrorealism. Moreover, cases have been identified where LGIs hide violations of macrorealism \cite{Avis:2010jm} that are detected by a simple NSIT condition \cite{Kofler:2013hb}. The latter fails for totally mixed initial states, where a more involved NSIT condition is required \cite{Clemente:2015hv}. These fundamental differences between tests of local realism and macrorealism seem connected to the peculiar definition of macrorealism \cite{Maroney:2014ws-arxiv, Bacciagaluppi:2014ue}.

In this paper, we analyze the reasons for and the consequences of this difference. We show that the probability space spanned by quantum mechanics (QM) is of a higher dimension in an MR test than in an LR test, and we analyze the resulting structure of the probability polytope. We conclude that inequalities---excluding the pathological case of inequalities pairwise merging into equalities---are not suited to be sufficient conditions for MR, and form only weak necessary conditions. Fine's theorem \cite{Fine:1982ic}, therefore cannot be transferred to macrorealism (unless one uses potentially negative quasi-probabiltities \cite{Halliwell:2015ws-arxiv}). Our study thus identifies a striking difference between the mathematical structures of LR and MR. While current experimental tests of macrorealism overwhelmingly use Leggett-Garg inequalities, this difference explains why NSIT is better suited as a witness of non-classicality, i.e.\ why it is violated for a much larger range of parameters~\cite{Kofler:2013hb, Clemente:2015hv}.

Let us start by reviewing the structure of the LR polytope (\textsf{LR}), as described in refs.\ \cite{Pironio:2005jb, Pironio:2014bx, Brunner:2014kr}. Consider an LR test between $n \geq 2$ parties $i \in \{1 \dots n\}$. Each party can perform a measurement in one of $m \geq 2$ settings $s_i \in \{1 \dots m\}$. Each setting has the same number $\Delta \geq 2$ of possible outcomes $q_i \in \{1 \dots \Delta\}$, and, to allow for all possible types of correlations, it may measure a distinct property of the system. We can define probability distributions $p_{q_1 \dots q_n | s_1 \dots s_n}$ for obtaining outcomes $q_1 \dots q_n$, given the measurement settings $s_1 \dots s_n$. If a party $i$ chooses not to perform a measurement, the corresponding ``setting'' is labeled $s_i=0$, and there is only one ``outcome'' labeled $q_i=0$ (e.g.\ $p_{q_1,0|s_1,0}$ when only the first party performs a measurement). We leave out final zeros, e.g.\ $p_{q_1 \dots q_i, 0 \dots 0 | s_1 \dots s_i, 0 \dots 0} = p_{q_1 \dots q_i | s_1 \dots s_i}$. Note that this convention differs from the literature for LR tests, where the case of no measurement is often left out \cite{Pironio:2005jb, Brunner:2014kr}, but simplifies the comparison between LR and MR tests. Each experiment is then completely described by $(m \Delta + 1)^n$ probability distributions; it can be seen as a point in a probability space $\mathbb R^{(m \Delta + 1)^n}$.

We now require normalization of the probabilities. There are $(m+1)^n$ linearly independent normalization conditions, as each probability only appears once:
\begin{equation}
  \forall s_1 \dots s_n\!: \sum_{q_1 \dots q_n} p_{q_1 \dots q_n | s_1 \dots s_n} = 1.
\end{equation}
Because of the special case of no measurements ($s_i = 0$), here (and in the following equations) we have abbreviated the notation of the summation: The possible values of $q_i$, in fact, depend on $s_i$.
The normalization conditions reduce the dimension of the probability space to
\begin{equation}\label{eq:dimP}
  (m \Delta + 1)^n - (m+1)^n.
\end{equation}
Furthermore, the positivity conditions
\begin{equation}
  \forall s_1 \dots s_n, q_1 \dots q_n\!: p_{q_1 \dots q_n | s_1 \dots s_n} \geq 0
\end{equation}
restrict the reachable space to a subspace with the same dimension, but they are delimited by flat hyperplanes. The resulting subspace is called the \emph{probability polytope} $\mathsf P$.

\begin{figure}[htb]
  \includegraphics[width=\columnwidth]{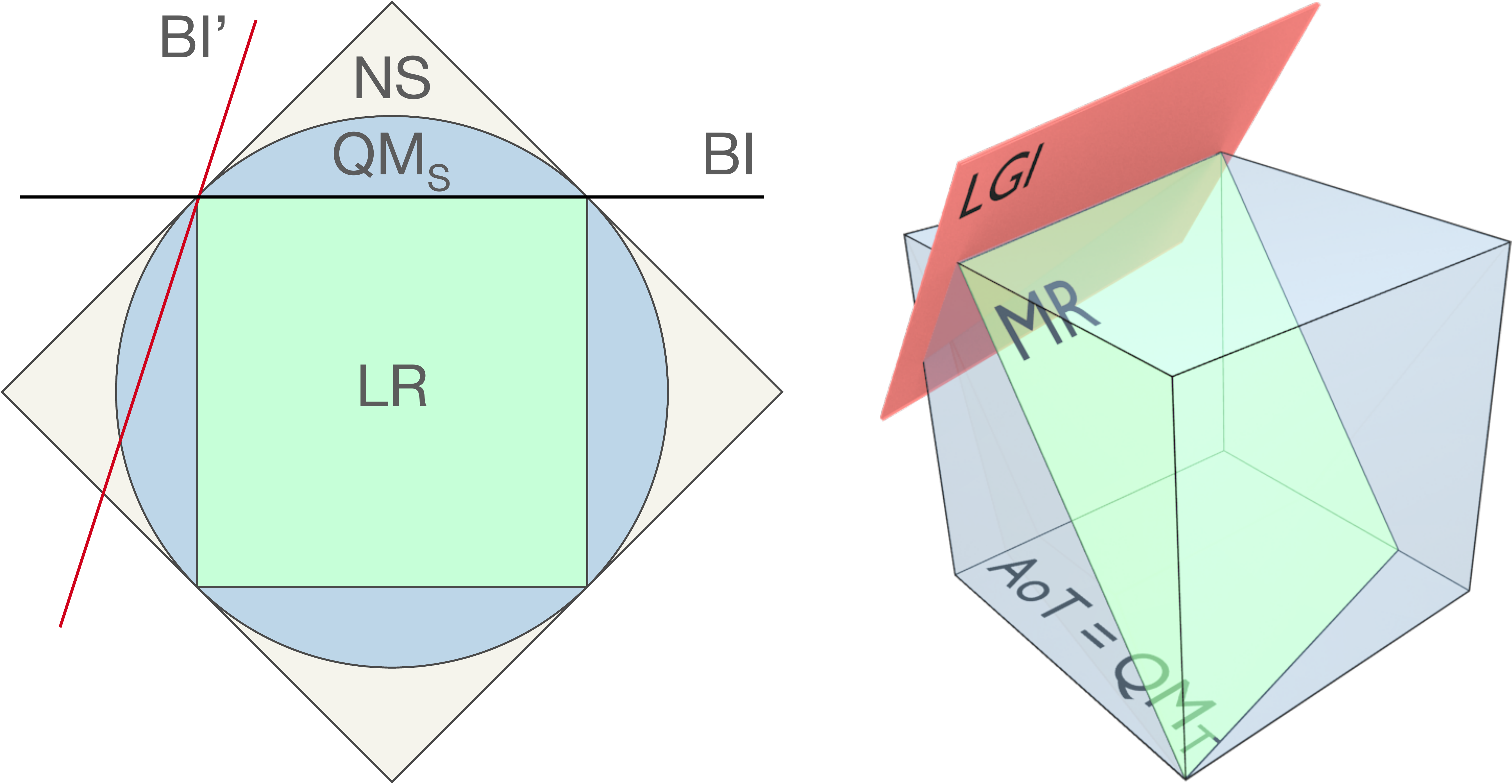}
  \caption{(Color online.)\label{fig:polytopes}
    \emph{Left:} A sketch of subspaces in an LR test \cite{Brunner:2014kr}. The no-signaling polytope (\textsf{NS}) contains the space of probability distributions realizable from spatially separated experiments in quantum mechanics ($\mathsf{QM_S}$), which contains the local realism polytope (\textsf{LR}). \textsf{LR} is delimited by Bell inequalities and the positivity conditions. \textsf{NS}, $\mathsf{QM_S}$, and \textsf{LR} have the same dimension. A Bell inequality (BI) is also sketched, delimiting \textsf{LR}. Another tight Bell inequality (BI') is less suited as a witness of non-LR behavior, and illustrates the role of Leggett-Garg inequalities in macrorealism tests.\\
    \emph{Right:} A sketch of polytopes in an MR test. The arrow of time polytope (\textsf{AoT}) is equal to the space of probability distributions realizable from temporally separated experiments in quantum mechanics ($\mathsf{QM_T}$), which contains the macrorealism polytope (\textsf{MR}). \textsf{MR} is a polytope of lower dimension, located fully within the $\mathsf{QM_T}$ subspace and solely delimited by positivity constraints. Since each probability can easily be minimized or maximized individually, \textsf{MR} reaches all facets of \textsf{AoT}. A Leggett-Garg inequality (LGI) is also sketched; it is a hyperplane of dimension $\dim \mathsf{QM_T}-1$, which, in general, is much larger than $\dim \mathsf{MR}$. Note that the LGI can only touch \textsf{MR} (i.e.\ be tight) at the boundary of the positivity constraints.
  }
\end{figure}

In an LR test with space-like separated parties, special relativity prohibits signaling from every party to any other,
\begin{equation}\label{eq:ns-conditions}
\begin{split}
  & \forall i, q_1 \dots q_{i-1}, q_{i+1} \dots q_n, s_1 \dots s_n, s_i \neq 0 \!:\\
  & p_{q_1 \dots q_{i-1}, 0, q_{i+1} \dots q_n | s_1 \dots s_{i-1}, 0, s_{i+1} \dots s_n} = \sum_{q_i=1}^\Delta p_{q_1 \dots q_n| s_1 \dots s_n}.
\end{split}
\end{equation}
These \emph{no-signaling} (NS) conditions restrict the probability polytope to a NS polytope (\textsf{NS}) of lower dimension. Taking their linear dependence, both amongst each other and with the normalization conditions, into account, we arrive at dimension \cite{Pironio:2005jb}
\begin{equation}
  \dim \mathsf{NS} = [m(\Delta-1)+1]^n - 1.
\end{equation}
Since quantum mechanics obeys NS, and due to Tsirelson bounds \cite{Cirelson:1980fp}, the space of probability distributions from spatially separated experiments implementable in quantum mechanics, $\mathsf{QM_S}$, is located strictly within the NS polytope. Furthermore, the space of local realistic probability distributions, \textsf{LR}, is a strict subspace of $\mathsf{QM_S}$. It is delimited by Bell inequalities (e.g. the CH/CHSH inequalities for $n = m = \Delta = 2$) and positivity conditions, and therefore forms a polytope within $\mathsf{QM_S}$ \cite{Fine:1982ic, Pironio:2005jb}. In summary, we have $\mathsf P \supset \mathsf{NS} \supset \mathsf{QM_S} \supset \mathsf{LR}$, with $\dim \mathsf P > \dim \mathsf{NS} = \dim \mathsf{QM_S} = \dim \mathsf{LR}$. The structure of the \textsf{NS}, $\mathsf{QM_S}$ and \textsf{LR} spaces is sketched on the left of \cref{fig:polytopes}.

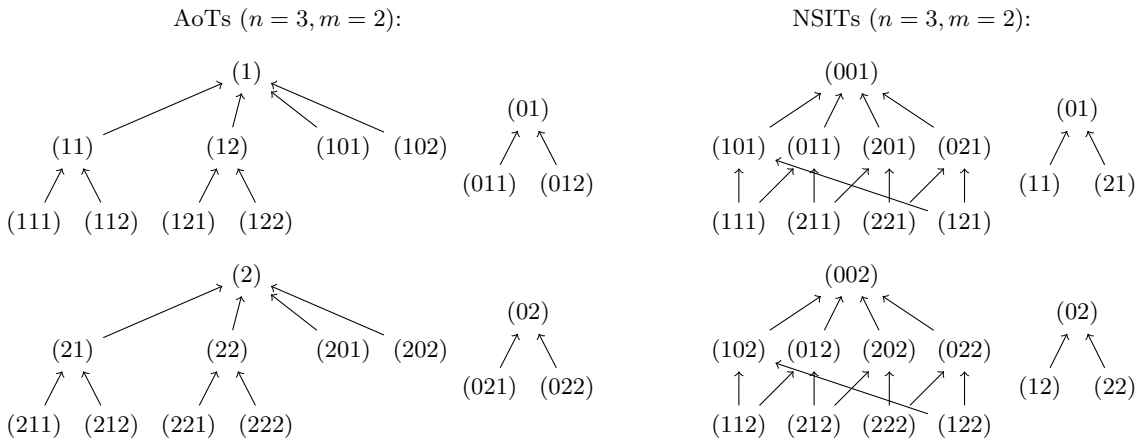
\begin{figure*}[ht]
  \begin{tikzpicture}[<-,
                      level distance = 1cm,
                      edge from parent path={(\tikzparentnode) -- (\tikzchildnode)}
                     ]
    \node at (-2.2, 0.7) {AoTs ($n = 3, m = 2$):};
    \node[frontier/.style={distance from root=2cm}] at (-3, -1) {\Tree
      [.(1)
        [.(11) (111) (112) ]
        [.(12) (121) (122) ]
        (101)
        (102)
      ]
    };
    \node[frontier/.style={distance from root=1cm}] at (1, -1) {\Tree
      [.(01) (011) (012) ]
    };
    \node[frontier/.style={distance from root=2cm}] at (-3, -3.7) {\Tree
      [.(2)
        [.(21) (211) (212) ]
        [.(22) (221) (222) ]
        (201)
        (202)
      ]
    };
    \node[frontier/.style={distance from root=1cm}] at (1, -3.7) {\Tree
      [.(02) (021) (022) ]
    };

    \tikzstyle{level 1}=[sibling distance = 1cm]
    \node at (6.1, 0.7) {NSITs ($n = 3, m = 2$):};
    \node at (5.3, 0) {(001)}
      child { node (nsit101) {(101)}
        child { node (nsit111) {(111)} }
      }
      child { node (nsit011) {(011)}
        child { node (nsit211) {(211)} }
      }
      child { node (nsit201) {(201)}
        child { node (nsit221) {(221)} }
      }
      child { node (nsit021) {(021)}
        child { node (nsit121) {(121)} }
      }
    ;
    \draw (nsit101) -- (nsit121);
    \draw (nsit011) -- (nsit111);
    \draw (nsit201) -- (nsit211);
    \draw (nsit021) -- (nsit221);
    \node at (5.3, -2.7) {(002)}
      child { node (nsit102) {(102)}
        child { node (nsit112) {(112)} }
      }
      child { node (nsit012) {(012)}
        child { node (nsit212) {(212)} }
      }
      child { node (nsit202) {(202)}
        child { node (nsit222) {(222)} }
      }
      child { node (nsit022) {(022)}
        child { node (nsit122) {(122)} }
      }
    ;
    \draw (nsit102) -- (nsit122);
    \draw (nsit012) -- (nsit112);
    \draw (nsit202) -- (nsit212);
    \draw (nsit022) -- (nsit222);
    \node at (8.3, -0.5) {(01)}
      child { node {(11)} }
      child { node {(21)} }
    ;
    \node at (8.3, -3.2) {(02)}
      child { node {(12)} }
      child { node {(22)} }
    ;
  \end{tikzpicture}
  \caption{\label{fig:independence}
    Arrow of time (AoT) and no-signaling in time (NSIT) conditions relating different outcome probability distributions for the case $n=3$ measurement times and $m=2$ possible settings. The notation $(xyz)$ refers to distributions with settings $s_1=x, s_2=y, s_3=z$. The arrows denote the process of marginalization: e.g.\ the AoT condition $p_{q_1|s_1=x} = \sum_{q_2} p_{q_1, q_2 | s_1=x, s_2=y}$ is denoted by $(x) \leftarrow (xy)$, and the NSIT condition $p_{q_2|s_2=y} = \sum_{q_1} p_{q_1, q_2 | s_1=x, s_2=y}$ is denoted by $(y) \leftarrow (xy)$. It can easily be seen that the AoT conditions are linearly independent, since they cannot form loops. Adding more measurement times (adding further rows), or adding more settings (broadening the trees) does not change their independence. In contrast, the NSIT conditions are not linearly independent, and thus form loops. Note that marginalizing only over a single measurement is sufficient, as simultaneous marginalizations follow from individual ones, and hence they are always linearly dependent.
  }
\end{figure*}

In a test of MR, temporal correlations take the role of an LR test's spatial correlations. Instead of spatially separated measurements on $n$ systems by different observers, a single observer performs $n$ sequential (macroscopically distinct) measurements on one and the same system. Again, each measurement is either skipped (``0'') or performed in one of $m \geq 1$ \footnote{In contrast to LR tests, where $m \geq 2$ is required to observe quantum violations, $m=1$ allows for violations of MR, and is in fact the most considered case in the literature.} settings, with $\Delta$ possible outcomes each. With this one-to-one correspondence, the resulting probability polytope $\mathsf P$ in the space $\mathbb R^{(m \Delta+1)^n-(m+1)^n}$ is identical to the one in the Bell scenario. However, without further physical assumptions, no-signaling in temporally separated experiments is only a requirement in one direction: While past measurements can affect the future, causality demands that future measurements cannot affect the past. This assumption is captured by the \emph{arrow of time} (AoT) conditions:
\begin{equation}\label{eq:aot-conditions}
  \begin{split}
    & \forall i \geq 2 \!: \forall q_1 \dots q_{i-1}, s_1 \dots s_{i-1} \text{ with } \Sigma_{j=1}^{i-1} s_j \neq 0, s_i \neq 0 \!:\\
    & p_{q_1 \dots q_{i-1} | s_1 \dots s_{i-1}} = \sum_{q_i=1}^\Delta p_{q_1 \dots q_i| s_1 \dots s_i}.
  \end{split}
\end{equation}
Counting the number of equalities in \cref{eq:aot-conditions} shows that their number is
\begin{equation}\label{eq:nAoT}
  \sum_{i=2}^{n} [(m \Delta + 1)^{i-1} - 1] m = \frac{(m \Delta + 1)^n - n m \Delta - 1}{\Delta},
\end{equation}
where the first factor in the sum counts the setting and outcome combinations for times $1 \dots i-1$, excluding the choice of all $s_i=0$, and the second factor the number of settings at time $i$. All listed conditions are linearly independent due to their hierarchical construction, see \cref{fig:independence}. However, a number of the normalization conditions for the marginal distributions, already subtracted in \cref{eq:dimP}, are not linearly independent from AoT, and thus become obsolete. Their number is obtained by counting the different settings in \cref{eq:aot-conditions}:
\begin{equation}\label{eq:nNormAoT}
  \sum_{i=2}^{n} [(m+1)^{i-1} - 1] m = (m+1)^n - nm - 1.
\end{equation}
The remaining normalization conditions are the ones for probability distributions with just one measurement and for the ``0-distribution''; there are $n m + 1$ such distributions.
Taking \cref{eq:dimP}, subtracting \cref{eq:nAoT} and adding \cref{eq:nNormAoT}, we conclude that the AoT conditions restrict the probability polytope to an AoT polytope (\textsf{AoT}) of dimension
\begin{equation}\label{eq:dimAoT}
  \dim \mathsf{AoT} = \frac{[(m \Delta + 1)^n - 1] (\Delta - 1)}{\Delta}.
\end{equation}
By simple extension of the proof in ref.\ \cite{Clemente:2015hv}, the set of all \emph{no-signaling in time} (NSIT) conditions,
\begin{equation}\label{eq:nsit-conditions}
  \begin{split}
    & \forall i \!<\! n, q_1 \dots q_{i-1}, q_{i+1} \dots q_n, s_1 \dots s_n, \Sigma_{j>i} s_j \!\neq\! 0, s_i \!\neq\! 0\!:\\
    & p_{q_1 \dots q_{i-1}, 0, q_{i+1} \dots q_n | s_1 \dots s_{i-1}, 0, s_{i+1} \dots s_n} = \sum_{q_i=1}^\Delta p_{q_1 \dots q_n| s_1 \dots s_n},
  \end{split}
\end{equation}
is, together with AoT, necessary and sufficient for macrorealism. To get from \textsf{AoT} to the macrorealism polytope, \textsf{MR}, we therefore require a linearly independent subset of these conditions. However, since the AoT conditions from \cref{eq:aot-conditions} plus the NSIT conditions from \cref{eq:nsit-conditions} are equivalent to the NS conditions from \cref{eq:ns-conditions}, we arrive at \textsf{MR} with the same dimension as the LR polytope:
\begin{equation}
  \dim \mathsf{MR} = \dim \mathsf{LR} = [m(\Delta-1)+1]^n - 1.
\end{equation}

We are left with the question of how the space of probability distributions realizable from temporally separated experiments in quantum mechanics, $\mathsf{QM_T}$, relates to \textsf{AoT}. Fritz has shown in ref.~\cite{Fritz:2010ba} that $\mathsf{QM_T} = \mathsf{AoT}$ for $n = m = \Delta = 2$, if we allow for positive-operator valued measurements (POVMs). Let us now generalize his proof to arbitrary $n, m, \Delta$. We do so by constructing a quantum experiment that produces all possible probability distributions which are allowed by AoT.

\newcommand{\vsym}[1]{\rotatebox[origin=c]{-90}{$#1$}}
\begin{table*}[tb]
  \renewcommand*{\arraystretch}{1.5}
  \begin{ruledtabular}
    \begin{tabular}{lcccc}
                                                & LR test                &                                                        & MR test \\\hline
      Number of unnormalized distributions      &                        & \hspace{-5em} $(m \Delta+1)^n$ \hspace{-5em}           & \\
      $\dim \mathsf P$                          &                        & \hspace{-5em} $(m \Delta+1)^n - (m+1)^n$ \hspace{-5em} & \\
      $\dim \mathsf{QM_S},~ \dim \mathsf{QM_T}$ & $[m (\Delta-1)+1]^n-1$ & $<$                                                    & $[(m \Delta + 1)^n - 1] (\Delta - 1)/ \Delta$ \\
      $\dim \mathsf{LR},~ \dim \mathsf{MR}$     &                        & \hspace{-5em} $[m (\Delta-1)+1]^n-1$ \hspace{-5em}     &
    \end{tabular}
  \end{ruledtabular}
  \caption{\label{tbl:dimensions}
    Dimensions of the probability space $\mathsf P$ and its subspaces reachable by spatially separated ($\mathsf{QM_S}$) or temporally separated ($\mathsf{QM_T}$) experiments in quantum mechanics, local realism (\textsf{LR}), and macrorealism (\textsf{MR}). There are $n$ spatially or temporally separated measurements with $m$ settings and $\Delta$ outcomes each.
  }
\end{table*}

Consider a quantum system of dimension $(m \Delta + 1)^n$, with states enumerated as $\ket{q_1 \dots q_n; s_1 \dots s_n}$. As with the probability distributions, final zeros may be omitted. The initial state of the system is $\ket{0 \dots 0; 0 \dots 0}$. Now, $n$ POVMs are performed on the system. The measurements are chosen such that depending on their setting and outcome they take the system to the corresponding state: Performing a measurement on a system in state $\ket{q_1 \dots q_{i-1}; s_1 \dots s_{i-1}}$ with setting $s_i$ and obtaining outcome $q_i$ should leave the system in state $\ket{q_1 \dots q_i; s_1 \dots s_i}$. This is accomplished by choosing Kraus operators for the $i$-th measurement in basis $s_i$ for outcome $q_i$ as
\begin{equation}\label{eq:kraus-construction}
\begin{split}
  K^{(i)}_{s_i, q_i} = &\sum_{s_1 \dots s_{i-1}, q_1 \dots q_{i-1}} \sqrt{r_{q_i | q_1 \dots q_{i-1}, s_1 \dots s_i}} \\
  &\times \ketbra{q_1 \dots q_i; s_1 \dots s_i}{q_1 \dots q_{i-1}; s_1 \dots s_{i-1}} \\
  + \sum_{\substack{s_1 \dots s_n \\ q_1 \dots q_n \\ \Sigma_{j=i}^{n} s_j \neq 0}} &\frac{1}{\sqrt{\Delta}} \ketbra{q_1 \dots q_n; s_1 \dots s_n}{{q_1 \dots q_n; s_1 \dots s_n}}.
\end{split}
\end{equation}
For $i = 1$, the first sum in \cref{eq:kraus-construction} reduces to the single term $\sqrt{p_{q_1 | s_1}} \ketbra{q_1; s_1}{0 \dots 0; 0 \dots 0}$, while the second sum remains unchanged.
The second sum in \cref{eq:kraus-construction} is necessary for the completeness relation $\sum_{q_i} (K^{(i)}_{s_i, q_i})^\dagger K^{(i)}_{s_i, q_i} = \mathbbm 1$. The above definitions also work for $s_i = 0$, where $r_{q_i = 0|q_1 \dots q_{i-1}, s_1 \dots s_{i-1}, s_i = 0} = 1$, and $(K^{(i)}_{s_i, q_i})^\dagger K^{(i)}_{s_i, q_i} = \mathbbm 1$.
The conditional probabilities $r$ in \cref{eq:kraus-construction} can be obtained from the probabilities $p$ using the assumption of AoT:
\begin{equation}
  r_{q_i | q_1 \dots q_{i-1}, s_1 \dots s_i} = \frac{p_{q_1 \dots q_i | s_1 \dots s_i}}{p_{q_1 \dots q_{i-1} | s_1 \dots s_{i-1}}}.
\end{equation}
This construction gives a recipe to obtain any point in the AoT probability space in a quantum experiment. We have therefore shown that $\mathsf{AoT} = \mathsf{QM_T}$ for any choice of $n, m, \Delta$.

Note that the probability distributions constructed above can also be achieved by a purely classical stochastic model, albeit with invasive measurements. Such an experiment would therefore not convince a macrorealist to give up their world view. For that to happen, an experiment needs to properly address the clumsiness loophole \cite{Leggett:1985bl, Wilde:2011ip, Moreira:2015iz}. The relevant methods previously established for the LGI can also be applied to NSIT-based experiments \cite{Knee:2016wd-arxiv}.

Since \textsf{AoT} is a polytope, $\mathsf{QM_T}$ with POVMs is also a polytope, and no non-trivial Tsirelson-like bounds exist. If, on the other hand, we only allowed projective measurements, we would have $\mathsf{QM_T} \subset \mathsf{AoT}$ with non-trivial Tsirelson-like bounds, as shown in ref.~\cite{Fritz:2010ba}. In this case, $\mathsf{QM_T}$ would not be a polytope. It is easy to see that QM with projectors is unable to reproduce some probability distributions: $n = 2, m=1, \Delta = 2$, $p_{11|11} = 1, p_{01|01} = 0$ fulfills AoT but cannot be constructed in projective quantum mechanics, since the initial state must be an eigenstate of the first measurement. Here we consider the general case of POVMs.

In summary, we have
\begin{equation}
  \begin{matrix}
    \mathsf P & \supset & \mathsf{NS}    & \supset & \mathsf{QM_S}  & \supset & \mathsf{LR} \\
    \vsym{=}  &         & \vsym{\subset} &         & \vsym{\subset} &         & \vsym{\subset} \\
    \mathsf P & \supset & \mathsf{AoT}   & =       & \mathsf{QM_T}  & \supset & \mathsf{MR}
  \end{matrix},
\end{equation}
with $\mathsf{NS} = \mathsf{MR}$, and dimensions
\begin{equation}
  \begin{matrix}
    \dim \mathsf P & > & \dim \mathsf{NS}  &                       = & \dim \mathsf{QM_S} & = & \dim \mathsf{LR} \\
    \vsym{=}       &   & \vsym{<}          &                         & \vsym{<}           &   & \vsym{=} \\
    \dim \mathsf P & > & \dim \mathsf{AoT} &                       = & \dim \mathsf{QM_T} & > & \dim \mathsf{MR}
  \end{matrix}.
\end{equation}
The structure of \textsf{AoT}, $\mathsf{QM_T}$ and \textsf{MR} within $\mathsf P$ is sketched on the right of \cref{fig:polytopes}, the dimensions of all mentioned subspaces are printed in \cref{tbl:dimensions}.

Finally, let us compare the characteristics of quantum mechanics in LR and MR tests. Trivially, QM fulfills NS between spatially separated measurements, and AoT between temporally separated measurements \footnote{%
  To show that QM fulfills NS, we consider a setup with only two parties, 1 and 2, performing measurements with POVM elements $\hat M_{q_1,s_1}^\dagger \hat M_{q_1,s_1}$ and $\hat M_{q_2,s_2}^\dagger \hat M_{q_2,s_2}$, respectively, on a two-particle state $\hat \rho_{12}$. We then calculate
  $
  \sum_{q_2} p_{q_1 q_2|s_1 s_2}
  = \sum_{q_2} \tr[(\hat M_{q_1,s_1}^\dagger \hat M_{q_1,s_1}) \!\otimes\! (\hat M_{q_2,s_2}^\dagger \hat M_{q_2,s_2})\,\hat \rho_{12}]§
  = \tr[(\hat M_{q_1,s_1}^\dagger \hat M_{q_1,s_1} \!\otimes\! \mathbbm{1}_2) \,\hat \rho_{12}]
  = \tr_1[\hat M_{q_1,s_1}^\dagger \hat M_{q_1,s_1} \tr_2(\hat \rho_{12}) ]
  = \tr_1 [\hat M_{q_1,s_1}^\dagger \hat M_{q_1,s_1} \hat \rho_{1}]
  = p_{q_1|s_1}
  $.
  To show that QM fulfills AoT, we consider a setup where $\hat M_{q_1,s_1}^\dagger \hat M_{q_1,s_1}$ are measured at time $1$ on state $\hat \rho_1$, and $\hat M_{q_2,s_2}^\dagger \hat M_{q_2,s_2}$ are measured at time $2$. We then have
  $
  \sum_{q_2} p_{q_1 q_2|s_1 s_2}
  = \sum_{q_2} \tr[\hat M_{q_1,s_1}^\dagger \hat M_{q_1,s_1} \hat \rho_1] \tr[\hat M_{q_2,s_2}^\dagger \hat M_{q_2,s_2} \hat \rho_2^{q_1,s_1}]
  = \tr[\hat M_{q_1,s_1}^\dagger \hat M_{q_1,s_1} \hat \rho_1]
  = p_{q_1|s_1}
  $, where $\hat \rho_2^{q_1,s_1}$ is the state after measurement of $s_1$ at time 1 with outcome $q_1$, evolved to time 2. The proofs for more parties or more measurement times follow straightforwardly.
}. While $\mathsf{QM_S}$ and \textsf{LR} have the same dimension and are separated by Bell inequalities, $\mathsf{QM_T}$ and \textsf{MR} span subspaces with different dimensions. Inequalities can never reduce the dimension of the probability space, since they act as a hyperplane separating the fulfilling from the violating volume of probability distributions. We conclude that no combination of (Leggett-Garg) inequalities can be sufficient for macrorealism.

The observation that inequalities cannot be sufficient for macrorealism, and the differences in the structure of the probability space shown above, present fundamental discrepancies between LR and MR. Fine's observation \cite{Fine:1982ic} that Bell inequalities are necessary and sufficient for LR can therefore not be transferred to the case of LGIs and MR. More precisely, Fine's proof uses the implicit assumption of NS, which is obeyed by all reasonable physical theories, including QM. However, the temporal analogue to NS is the conjunction of AoT and NSIT, where AoT is obeyed by all reasonable physical theories, while NSIT is violated in QM. Therefore,
\begin{alignat}{3}
  \text{BIs} &~\substack{\Leftarrow\\\nRightarrow}~ \text{LR} && \Leftrightarrow \text{NS} \land \text{BIs} \\
  \text{LGIs} &~\substack{\Leftarrow\\\nRightarrow}~ \text{MR} && \Leftrightarrow \text{AoT} \land \text{NSIT} ~\substack{\nLeftarrow\\\Rightarrow}~ \text{AoT} \land \text{LGIs},
\end{alignat}
where ``BIs'' and ``LGIs'' denote the sets of all Bell and Leggett-Garg inequalities, respectively.

Moreover, since \textsf{MR} is a polytope with smaller dimension than $\mathsf{QM_T}$, LGIs can only touch \textsf{MR} (i.e.\ be \emph{tight}) at one facet, i.e.\ a positivity constraint, as sketched in \cref{fig:polytopes} on the right. A comparable Bell inequality, sketched in \cref{fig:polytopes} on the left as BI', clearly illustrates the limitations resulting from this requirement. In an experimental test of MR, using a LGI therefore needlessly restricts the parameter space where violations can be found. The favorable experimental feasibility of NSIT is demonstrated by the theoretical analyses of refs.~\cite{Kofler:2013hb, Clemente:2015hv}, as well as the recent experiment of ref.~\cite{Knee:2016wd-arxiv}. Note also the mathematical simplicity of the NSIT conditions when compared to the LGI. We conclude that for further theoretical studies and future experiments it might be advantageous to eschew the LGIs and rather use NSIT.

\begin{acknowledgments}
  We acknowledge support from the EU Integrated Project SIQS.
\end{acknowledgments}

\bibliography{LR-MR_polytopes,arxiv}

\end{document}